\documentclass[useAMS,usenatbib]{mn2e}
\usepackage{psfig}
\usepackage{amsmath}

\begin{document}

\title[Magnetic fields in AGNs and microquasars]
{Magnetic fields in AGNs and microquasars}
\author[A.F.Zakharov, N.S.Kardashev, 
V.N.Lukash and S.V.Repin]
{A.F.Zakharov$^{1,2}$,\thanks{E-mail: zakharov@vitep1.itep.ru (AFZ),
kardashev@asc.rssi.ru (NSK),
vladimir@lukash.asc.rssi.ru (VNL), repin@mx.iki.rssi.ru (SVR)}
 N.S.Kardashev$^{2}$, 
V.N.Lukash$^{2}$  and S.V.Repin$^{3}$
\\
$^1$ Russian Scientific Centre --
Institute of Theoretical and Experimental Physics, 117259, Moscow,\\
$^2$ 
Astro Space Centre of Lebedev Physics Institute, Moscow\\
$^3$ Space Research Institute of Russian Academy of Sciences, Moscow\\
}

\maketitle

\begin{abstract}
 Observations of AGNs and microquasars by ASCA, RXTE, Chandra
and XMM-Newton indicate the existence of wide X-ray emission
lines of heavy ionized elements in their spectra.
The emission can arise in the inner parts of accretion discs
where the effects of General Relativity (GR) must be counted,
moreover such effects can dominate.
We describe a procedure to estimate an upper limit of 
the magnetic fields in the regions where X-ray photons are emitted.
We simulate typical profiles
of the iron $K_\alpha$ line in the presence of a magnetic field
and compare them with observational data. 
As an illustration we find $H < 10^{10} - 10^{11}$~Gs for
Seyfert galaxy MCG--6--30--15.
Using the perspective facilities of measurement devices (e.g.
Constellation-X mission) a better resolution of the blue peak
structure of iron $K_\alpha$ line will allow to find the
value of the magnetic fields if the latter are high enough.

\end{abstract}


\section{Introduction}

   Recent ASCA, RXTE, Chandra and XMM-Newton observations
of Seyfert galaxies demonstrated the existence of the wide
iron $K_\alpha$ line (6.4~keV) in their spectra along with
a number of other weaker lines
(Ne~X, Si~XIII,XIV, S~XIV-XVI, Ar~XVII,XVIII, Ca~XIX, etc.)
(see for example,
\citet{fabian1,tanaka1,nandra1,nandra2,malizia,sambruna,
yacoob4,ogle1}).

     Magnetic fields play a key role in dynamics of accretion
discs and jet formation.
\cite{Bis74,Bis76} considered a scenario to generate
superstrong magnetic fields near black holes.
According to their results magnetic fields
near the marginally stable orbit could be about
$H \sim 10^{10} - 10^{11}$~Gs.
\citet{Kard95,Kard00,Kard01a,Kard01} considered
a generation of synchrotron radiation, acceleration of
$e^{+/-}$ pairs and cosmic rays in magnetospheres
of supermassive black holes. It is magnetic field, which
plays a key role in these models. Below, based on the analysis
of iron $K_\alpha$ line profile in the presence of a strong
magnetic field, we describe how to
detect the field itself or at least
obtain an upper limit of the magnetic field.

     For cases when the spectral resolution is good enough
the emission spectral line demonstrates typical two-peak
profile with the high "blue" peak, the low "red" peak
and the long "red" wing which drops gradually to the background
level (\cite{tanaka1,yaqoob2}). The Doppler line width
corresponds to the velocity of the matter motion of tens
of thousands kilometers per second \footnote{Note that
the measured line shape differs essentially from the
Doppler one.}, e.g. the maximum value is about
$v \approx 80000 - 100000$~km/s  for the galaxy
MCG--6--30--15 (\cite{tanaka1,Fabian02})
and $v \approx 48000$~km/s  for \mbox{MCG--5--23--16}
(\cite{krolik1}).
For both galaxies the line profiles are known rather well.
\cite{Fabian02} analyzed results of long-time observations
of MCG-6-30-15 galaxy using {\it XMM-Newton} and {\it BeppoSAX}.
The long monitoring confirmed the qualitative conclusions
about the features of the Fe $K_\alpha$ line, which were discovered
by ASCA satellite. \citet{yaqoob02} discussed the essential
importance of ASCA calibrating and the reliability of obtained
results. \cite{Lee02} compared ASCA results with RXTE and Chandra
observations for the MCG-6-30-15. \cite{Iwa99,Lee99,Shih02}
analyzed in detail the variability in continuum and in
Fe $K_\alpha$ line for the MCG-6-30-15 galaxy.

     The phenomena of the broad emission lines are supposed 
to be related with accreting matter around black holes.
\cite{Wilms01,Ball01,Mart02} proposed physical models
of accretion discs for the MCG-6-30-15 galaxy and their
influence on the Fe $K_\alpha$ line shape.
\cite{boller01} found the features of the spectral line
near  7~keV in Seyfert galaxies using data from
{\it XMM-Newton} satellite.
\cite{yaqoob01a} presented results of
Chandra HETG observations of Seyfert~I galaxies.
~\cite{Qing01} discussed possible identification
of binary massive black holes analyzing Fe $K_\alpha$ shape.
\cite{Ball02} estimated abundance of the iron using the
data of X-ray observations.
\cite{Popov01,Popov02} discussed
an influence of microlensing on the distortion
of spectral lines including Fe $K_\alpha$ line, that
can be significant in some cases.
\cite{matt02} analyzed an influence of Compton effect
on the Fe $K_\alpha$ shape for emitted and reflected
spectra. \cite{moral01} proposed a procedure to estimate 
the masses for supermassive black holes.
\cite{fabian99a} presented a possible scenario
for evolution of such supermassive black holes.

     A general status of black holes is described in
a number of papers (see, e.g. \citet{Liang98}
and references therein, \citet{Zak00,FN01}).
Since the matter motions indicate very high rotational
velocities, one can assume the $K_\alpha$ line emission arises
in the inner regions of accretion discs
at distances $\sim (1\div 3)~r_g$ from the black holes.
Let us recall that the innermost stable circular
for non-rotational black hole (which has the Schwarzschild metric)
is located at the distance $3\,r_g$ from the black hole
singularity.
Therefore, a rotation of black hole could be
the most essential factor.
A possibility to observe the matter motion in
so strong gravitational fields could give
a chance not only to check general relativity predictions
and simulate physical conditions in accretion discs,
but investigate also observational manifestations
of such astrophysical phenomena like
jets \citep{romanova1,romanova2},
some instabilities like Rossby waves \citep{Love99} and
gravitational radiation.

     Wide spectral lines are considered to be formed by 
radiation  emitted in the vicinity of black holes.  If there are
strong magnetic fields near black holes these lines are split
by the field into several components. This phenomenon is
discussed below. Such lines have been 
found in microquasars, GRBs and other similar objects 
\citep{Balu99,grein99,mira00,Mira02,Lazz01,Mart02a,Mira02a,
Miller02,zaman02}.

     Observations and theoretical interpretations of wide
X-ray lines (particularly, the iron $K_\alpha$ line) in AGNs 
are actively discussed in a number of papers
\citep{yaqoob1,wanders,sulentic1,sulentic2,paul,Bia02,Tur02,
Lev02,Lev02a}.
The results of numerical simulations in framework of different 
physical assumptions on the origin of the wide emissive iron 
$K_\alpha$ line in the nuclei of Seyfert galaxies are presented
in papers 
\citep{matt1,bromley,pariev2,pariev1,cui,bromley2,pariev3,
Ma00,Ma02,karas01}.                            
The results of Fe $K_\alpha$ line observations and their 
possible interpretation are summarized by \citet{fabian2}.

     To obtain an estimation of the magnetic field we simulate
the formation of the line profile for different values of
magnetic field. As a result we find the minimal $B$ value
at which the distortion of the line profile becomes
significant. We use here an approach, which is based on
numerical simulations of trajectories of the photons emitted
by a hot ring moving along a circular geodesics near
black hole, described earlier by
\cite{zakharov6,zakharov1,zakharov5,zak_rep1}.

\section{Magnetic fields in accretion discs}

One of the basic problems to understand the physics of quasars
and microquasars is the "central engine" in these systems,
in particular, a physical mechanism to accelerate charged 
particles and generate high energetic electromagnetic radiation 
near black holes. The construction of such "central engine"
without magnetic fields could  hardly ever be possible.
On the other hand magnetic fields give a possibility
to extract energy from rotational black holes via
Penrose process and Blandford -- Znajek mechanism,
as it was shown in hydrodynamical simulations
by \cite{Koide02,Koide02a}. The Blandford -- Znajek process 
could provide huge energy release in AGNs (for example, 
for MCG-6-30-15) and microquasars when the magnetic field
is strong enough \citep{Wilms01}.

    Physical aspects of generation and evolution
of magnetic fields were considered in a set
of reviews (e.g. \cite{Ass87,Giov01}).
A number of papers conclude that in the vicinity of
the marginally stable orbit the magnetic fields could be
high enough \citep{Bis74,Bis76,Krolik99}.

     \cite{Agol99} considered an influence of magnetic 
fields on an accretion rate near the marginally stable orbit 
and hence on the disc structure,
they found the appropriate changes of the emitting 
spectrum and solitary spectral lines.
\cite{vietri98} investigated the instabilities of 
accretion discs when the magnetic fields play 
an important role.

\section{Photon geodesics in the Kerr metric}

     Many astrophysical processes where large energy release
is observed, are supposed to be related with black holes.
Since a large fraction of astronomical objects, such as stars
and galaxies, exhibits proper rotation, then there are
no doubts that the black holes formed in their nuclei,
both stellar and supermassive, possess an intrinsic proper
rotation. It is known that stationary black holes
are described by the Kerr metric which has the following form
in geometrical units $(G = c = 1)$
and the Boyer--Lindquist coordinates $(t,r,\theta,\phi)$
\citep{wheeler,fieldtheory}:
\begin{eqnarray}
  ds^2 = - \left( 1 - \frac{2Mr}{\rho^2} \right) dt^2 +
         \frac{\rho^2}{\Delta} dr^2 + \rho^2 d\theta^2 +
\nonumber  \\
       \left(
               r^2 + a^2 + \frac{2Mra^2}{\rho^2}\sin^2\theta
         \right) \sin^2\theta\, d\phi^2 -
         \frac{4Mra}{\rho^2}\sin^2\theta\, d\phi dt,
                      \label{eq2}
\end{eqnarray}
where
$$
   \rho^2 = r^2 + a^2 \cos^2\theta,
$$
$$
   \Delta = r^2 - 2Mr + a^2.
$$
Constants $M$ and $a$ determine the black hole parameters:
$M$ is its mass, $a \in (0,M)$ is its specific
angular moment.

     The particle trajectories can be described by the standard
geodesic equations:
\begin{equation}
\frac{d^2 x^i}{d\lambda^2}+\Gamma^i_{kl}
\frac{dx^k}{d\lambda}
\frac{dx^l}{d\lambda}=0,
 \label{eq4}
\end{equation}
 where  $\Gamma^i_{kl}$ are the Christoffel symbols and
$\lambda$ is the affine parameter.
These equations can be simplified if we will use
the complete set of the first integrals which were found by
\citet{carter}:
$E=p_t$ is the particle energy at infinity, $L_z = p_\phi$ is 
the projection of its angular momentum on the rotation axis, 
$m=(p_ip^i)^{1/2}$ is the particle mass and
$Q$ is the Carter separation constant:
\begin{equation}
   Q = p_\theta^2 + \cos^2\theta
                    \left[a^2 \left(m^2 - E^2\right) +
                          L_z^2 / \sin^2\theta
                    \right].     \label{eq5}
\end{equation}
The equations of photon motion ($m=0$) can be reduced to the
following system of ordinary differential equations
\citep{zakh91,zakharov1}:
\begin{eqnarray}
   \frac{dt}{d\sigma}
                           & = &
      - a \left(a \sin^2\theta - \xi\right) +
      \frac{r^2 + a^2}{\Delta}
       \left(r^2 + a^2 - \xi a\right),
                        \label{eq6}                       \\
   \frac{dr}{d\sigma} & = & r_1,       \label{eq7}        \\
   \frac{dr_1}{d\sigma}    & = &
      2r^3 + \left(a^2 - \xi^2 - \eta\right) r +
      \left(a - \xi\right)^2 + \eta,                      \\
   \frac{d\theta}{d\sigma} & = & \theta_1,                \\
   \frac{d\theta_1}{d\sigma}
                           & = &
      \cos\theta \left(\frac{\xi^2}{\sin^3\theta} -
                       a^2 \sin\theta
                 \right),              \label{eq10}       \\
   \frac{d\phi}{d\sigma}   & = &
      - \left(a - \frac{\xi}{\sin^2\theta}\right) +
      \frac{a}{\Delta}
           \left(r^2 + a^2 - \xi a \right),
                     \label{eq11}
\end{eqnarray}
where $\sigma$ is an independent variable, 
$\eta = Q/M^2E^2$ and $\xi = L_z/ME$ are the Chan\-dra\-sekhar's
constants \citep{chandra} which should be derived from 
initial conditions in the disc plane;
$t$, $r$ and $a$ are here the appropriate dimensionless variables 
(in the mass units). The system (\ref{eq6})-(\ref{eq11}) has 
two first integrals,
\begin{eqnarray}
  \epsilon_1 & \equiv & r_1^2 - r^4 -
      \left(a^2 - \xi^2 - \eta\right) r^2 -    \nonumber \\
            &        &
      \phantom{.} -
      2\left[\left(a - \xi\right)^2 + \eta \right] r +
      a^2\eta = 0,                  \\
  \epsilon_2 & \equiv & \theta_1^2 - \eta - \cos^2\theta
      \left(a^2 - \frac{\xi^2}{\sin^2\theta}\right) = 0,
                    \label{eq13}
\end{eqnarray}
which can be used for the accuracy control of computation.

     The additional variables $r_1$ and $\theta_1$
reduce the Eqs. (\ref{eq7})--(\ref{eq10}) to a non-singular
form. Such kind representation allows to avoid the integration
difficulties which usually appears when the equations are
written in standard form \citep{wheeler,fieldtheory}
for $r$ and $\theta$ coordinates.

     A qualitative analysis of the geodesic equations
showed that types of photon motion
can drastically change with small changes of chosen geodesic
parameters \citep{zakh86,zakh89}. Therefore, the standard way
where there is one equation for each Boyer-Lindquist coordinate
(reducing to calculation of the elliptical integrals) can
lead to large numerical errors \citep{zakh91}. The integration 
of Eqs. (\ref{eq6})--(\ref{eq11}) allows to avoid such problem 
and realizes this process without essential numerical errors.

     Solving Eqs. (\ref{eq6})--(\ref{eq11}) for 
monochromatic quanta emitted by a hot ring rotating on 
circular geodesics at radius $r$ in the equatorial plane, 
we can obtain the ring spectrum $I_\nu(r,\theta_\infty)$
which is registered by a distant observer in the position 
characterized by the angle 
$\theta_\infty = \theta\big|_{r=\infty}$.
The numerical integration is performed using the Gear and Adams 
methods ~\citep{gear} and the standard package realized 
by~\citet{hindmarsh,petzold,hiebert}. We obtain the entire 
disc spectrum by summation of sharp ring spectra.

\section{Disc radiation model}

    To simulate the radiated spectrum it is necessary first 
to adopt some emission model. We assume that the source of
the emitting quanta is a narrow and thin disc (ring) rotating 
in the equatorial plane of a Kerr black hole.
For $a=0.01$ the
marginally stable orbit lies at $r_{ms} = 2.9836\,r_g$,
the difference from $3\,r_g$ ($r_g = 2M$) is not important for 
our analysis. We also assume that disc is opaque to radiation, 
so that a distant observer situated on one 
disc side cannot register the quanta emitted fron its another side.

    For the sake of computational simplicity we suggest that the
spectral line is monochromatic in its co-moving frame. To approve 
this assumption one can argue that even at $T = 10^8$~K the 
thermal width of the line
$$
    \frac{\delta f}{f} \sim 
     \frac{1}{c}\sqrt{\frac{kT}{m_{_{\rm Fe}}}} \approx 10^{-3}
$$
appears to be much less than the Doppler line width associated
with the disc rotation.

    Note that we do {\bf not} assume any particular model of the 
accreting disc. {\bf As an illustration} 
we determine the dependence of disc temperature on
the radial coordinate according to the
standard $\alpha$-disc model
\citep{shasun,shakura,LipShak}.
The radiation intensity is, as usually, proportional
to~$T^4$.

     The emission intensity of the ring at a given radius $r$
is proportional to the area of the ring.
The area of emitting ring of the width $dr$ differs in the
Schwarzschild metric from its classical expression
$dS = 2\pi rdr$  and should be replaced with
\begin{equation}
   dS = \frac{2\pi r\,dr}
             {\displaystyle\sqrt{1 - \frac{r_g}{r}\phantom{1}}}.
          \label{eq12}
\end{equation}

    Thus, the total flux density emitted by the disc and 
registered by distant observer is proportional to the integral
\begin{equation}
   J_\nu(\theta_\infty) = \int\limits_{r_{in}}^{r_{out}}
              I_\nu(r,\theta_\infty) T^4(r) dS,
          \label{eq14}
\end{equation}
where $J_\nu(r,\theta_\infty)$ is obtained from the solution 
of equations (\ref{eq6})--(\ref{eq11}),
$T(r)$ -- from the appropriate dependence for $\alpha$-disc and 
$dS$ -- from Eq.~(\ref{eq12}).

    The radiation pressure predominates in the innermost 
part of 
the disc ($a$) while the gas pressure -- in the 
middle ($b$). The boundary $r_{ab}$ between these two regions 
can be found for $\alpha$-disc from the following equation 
\citep{shasun}
\begin{equation}
    \frac{r_{ab}/(3r_g)}
            {\left(1 - \sqrt{3r_g/r_{ab}}\right)^{16/21}} =
     150(\alpha M/M_\odot)^{2/21}\dot m^{16/21},
          \label{eq15}               
\end{equation}
which we solve by an iteration procedure. 
In Eq.(\ref{eq15}) we have 
$\dot m = \dot M/\dot M_{cr}$, 
where $\dot M_{cr} = 3\cdot 10^{-8} M$/yr.
Thus, for $\alpha=0.2$, $M = 10^8M_\odot$ and
$\dot m = 0.1$ we have from Eq.(\ref{eq15})
$r_{ab} \approx 360\,r_g$. 

     For simulation we assume that
the emitting region lies as whole in the innermost region of
the disc (zone $a$). If this condition
is not satisfied, the profile of a spectral line becomes
extremely complicated, so that it appears difficult
to avoid the uncertainty in interpretation. Thus, we assume
that emission arises in the region from $r_{out} = 10\,r_g$ to
$r_{in} = 3\,r_g$ and the emission is monochromatic in the
co-moving frame. Set the frequency of this emission as a unity 
by convention. 

\section{Influence of a magnetic field on the distortion
         of the iron $K_\alpha$ line profile}

    The magnetic pressure at the inner edges of the accretion
discs and its correspondence with parameter $a$ in the frameworks
of disc accretion models is discussed 
by~\cite{krolik01}. However, the numerical value of magnetic
field is determined there from a model-dependent procedure, 
a number of parameters in which cannot be found explicitly from
observations.

    Here we consider the influence of a magnetic field on
the iron $K_\alpha$ line profile \footnote{We can also consider 
X-ray lines of other elements emitted by the area of accretion 
disc close to the marginally stable orbit; further we say only 
about iron $K_\alpha$ line for brevity} and show how one can 
determine the value of the magnetic field strength or at least
an upper limit.

    The profile of a monochromatic line
\citep{zak_rep1,zak_rep2,zak_rept} depends on the angular momentum
of a black hole, the position angle between the black hole axis 
and the distant observer position, the value of the radial
coordinate if the emitting region represents an infinitesimal
ring (or two radial coordinates for
outer and inner bounds of a wide disc).

    We assume that the emitting region is located in
an area of strong quasi-static magnetic field. This field
causes line splitting due to the standard Zeeman effect. 
There are three characteristic frequencies of the split
line that arise in the emission. The energy of central 
component $E_0$ remains unchanged, whereas two extra
components are shifted by 
$\pm \mu_B H$, where
$\mu_B=\dfrac{e \hbar}{2m_{\rm e}c}=9.273\cdot 10^{-21}$~erg/Gs is
the Bohr magneton. Therefore, in the presence of a
magnetic field we have three energy levels:
$E_0-\mu_B H,~ E_0$ and $E_0+\mu_B H$. For the iron
$K_\alpha$ line they are as follows:
$E_0=6.4 - 0.58 \dfrac{H}{10^{11}\,{\rm Gs}} $ keV,
$E_0=6.4$ keV and
$E_0=6.4 + 0.58 \dfrac{H}{10^{11}\,{\rm Gs}} $ keV.

   Let us discuss how the line profile changes when photons
are emitted in the co-moving frame with energy $E_0 (1+\epsilon)$,
but not with $E_0$. In that case the line profile can be
obtained from the original one by $1+\epsilon$ times
stretching along the x-axis which counts the energy.
The component with $E_0 (1-\epsilon)$ energy should be
$(1-\epsilon)$ times stretched, respectively. 
The intensity of different Zeeman components are approximately 
equal \citep{Fock76}. A composite line profile
can be found by summation the initial line with energy $E_0$ and
two other profiles, obtained by stretching this line along the
$x$-axis in $(1+\epsilon)$ and $(1-\epsilon)$ times
correspondingly. The line intensity depends
on the direction of the quantum escape with respect to the direction
of the magnetic field \citep{BLP89}. However, we neglect this weak
dependence (undoubtedly, the dependence can be counted and,
as a result, some details in the spectrum profile can be slightly
changed, but the qualitative picture, which we discuss,
remains unchanged).

\begin{figure}
  \psfig{figure=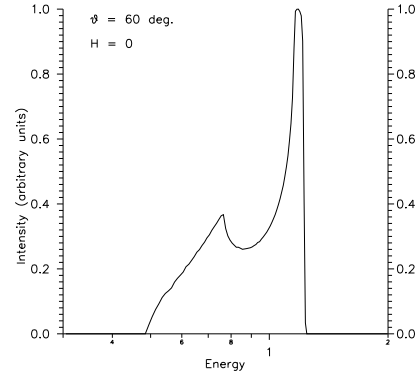,width=6cm}

  \psfig{figure=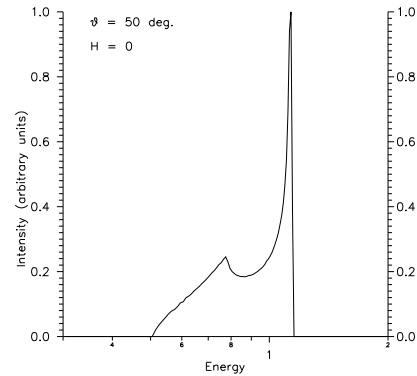,width=6cm}
  \caption{Profile of monochromatic spectral line, emitted
           by $\alpha$-disc in Schwarzschild metric for
           $r_{out} = 10\,r_g$, $r_{in} = 3\,r_g$ and
           inclination angles $\theta = 60^\circ$ (top panel)
           and $\theta = 50^\circ$ (bottom panel) with zero
           value of magnetic field. The line profile shown as
           it is registered by a distant observer.}
  \label{zeeman01}
\end{figure}

\begin{figure*}
  \psfig{figure=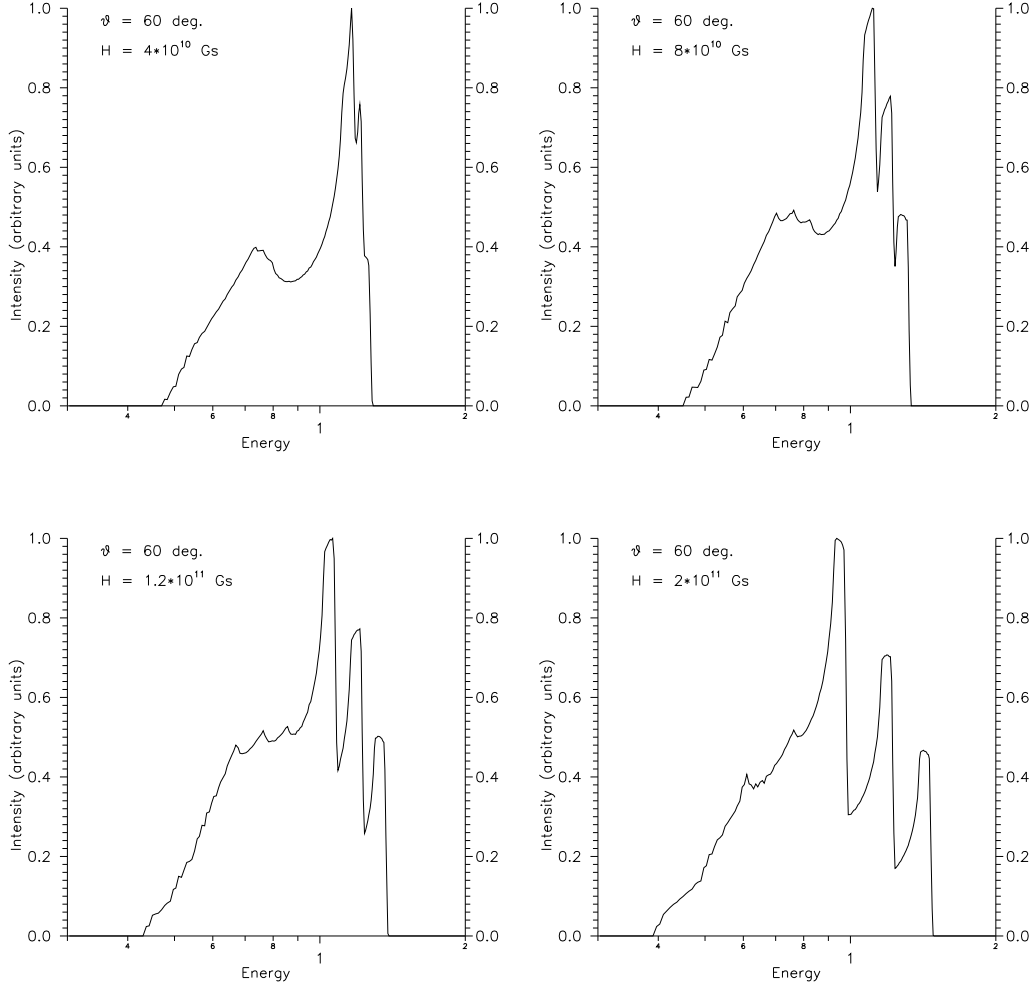,width=15cm}
  \caption{Distortions of the line profile from the top panel
           in Fig.~\ref{zeeman01}, arising due to a
           quasi-static magnetic field existing in the disc.
           The Zeeman effect leads to the appearance of two
           extra components with the energies higher and lower 
           than the basic one. The values of the magnetic
           field are shown at each panel.}
  \label{zeeman02}
\end{figure*}
\suppressfloats[b]

     Another indicator of the Zeeman effect is a significant
induction of the polarization of X-ray emission: the extra
lines possess a circular polarization (right and left, 
respectively, when they are observed along the field direction) 
whereas a linear polarization arises if the magnetic field is
perpendicular to the line of sight.\footnote{Note that another
possible polarization mechanisms in $\alpha$-disc were discussed 
by~\citet{Saz02}.} Despite of the fact that
the measurements of polarization of X-ray emission have not
been carried out yet, such experiments can 
be realized in the nearest future \citep{Cos01}.

\begin{figure*}
  \psfig{figure=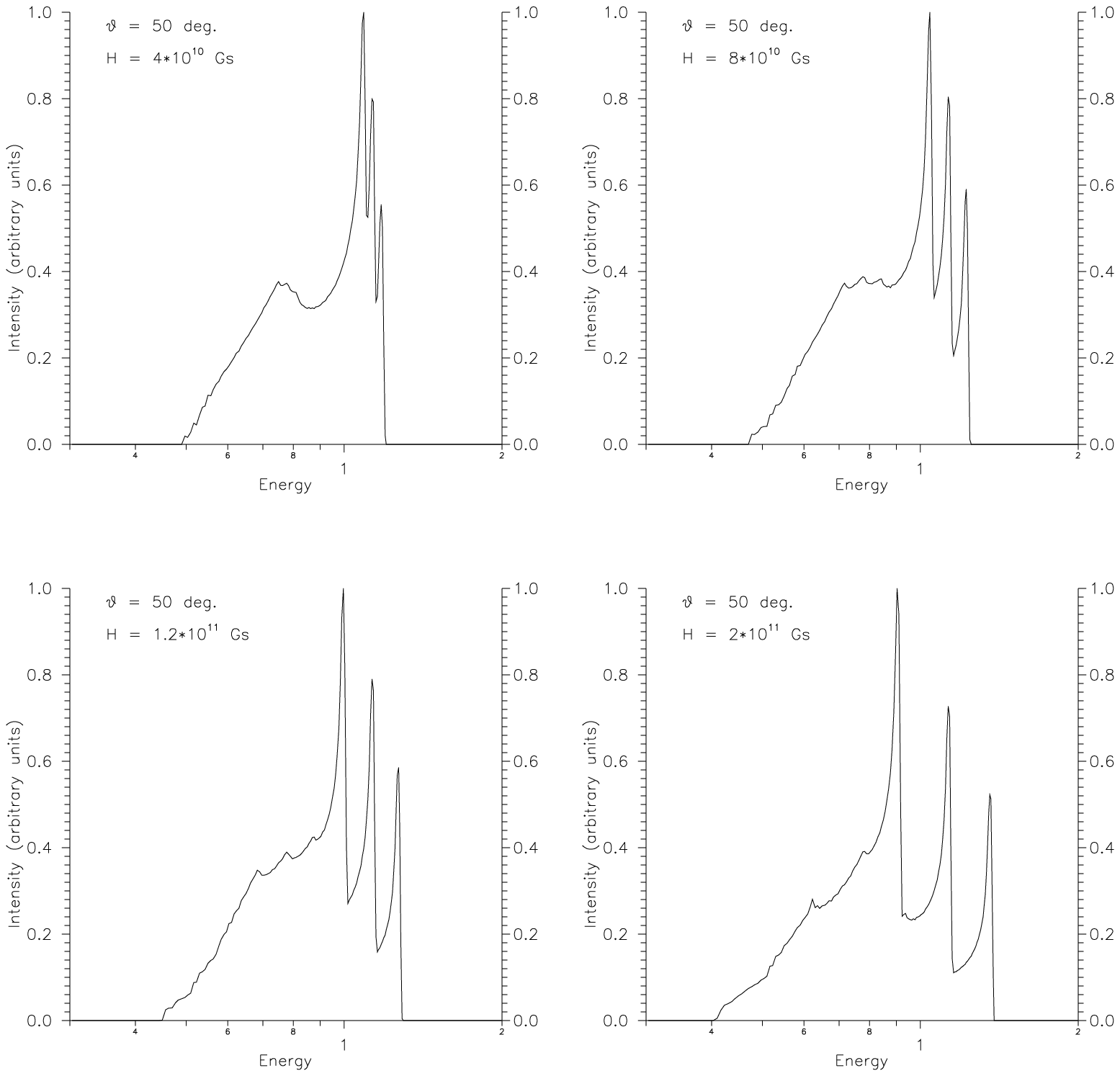,width=15cm}
  \caption{The same as in Fig.~\ref{zeeman02},
           but for $\theta = 50\degr$. The bottom panel on
           Fig.~\ref{zeeman01} demonstrates the same spectrum
           without magnetic field.}
  \label{zeeman03}
\end{figure*}

\begin{figure}
  \psfig{figure=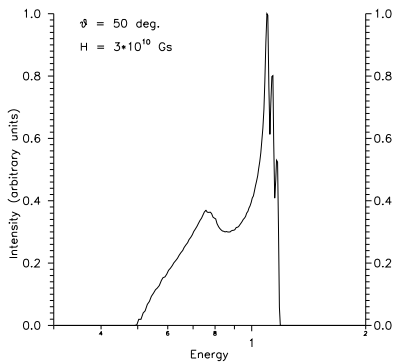,width=6cm}
  \caption{The same as in Fig.~\ref{zeeman03}, but for
           minimum value of magnetic field at which it is
           possible to observe the splitting of the line.}
  \label{zeeman04}
\end{figure}

\begin{figure*}
  \psfig{figure=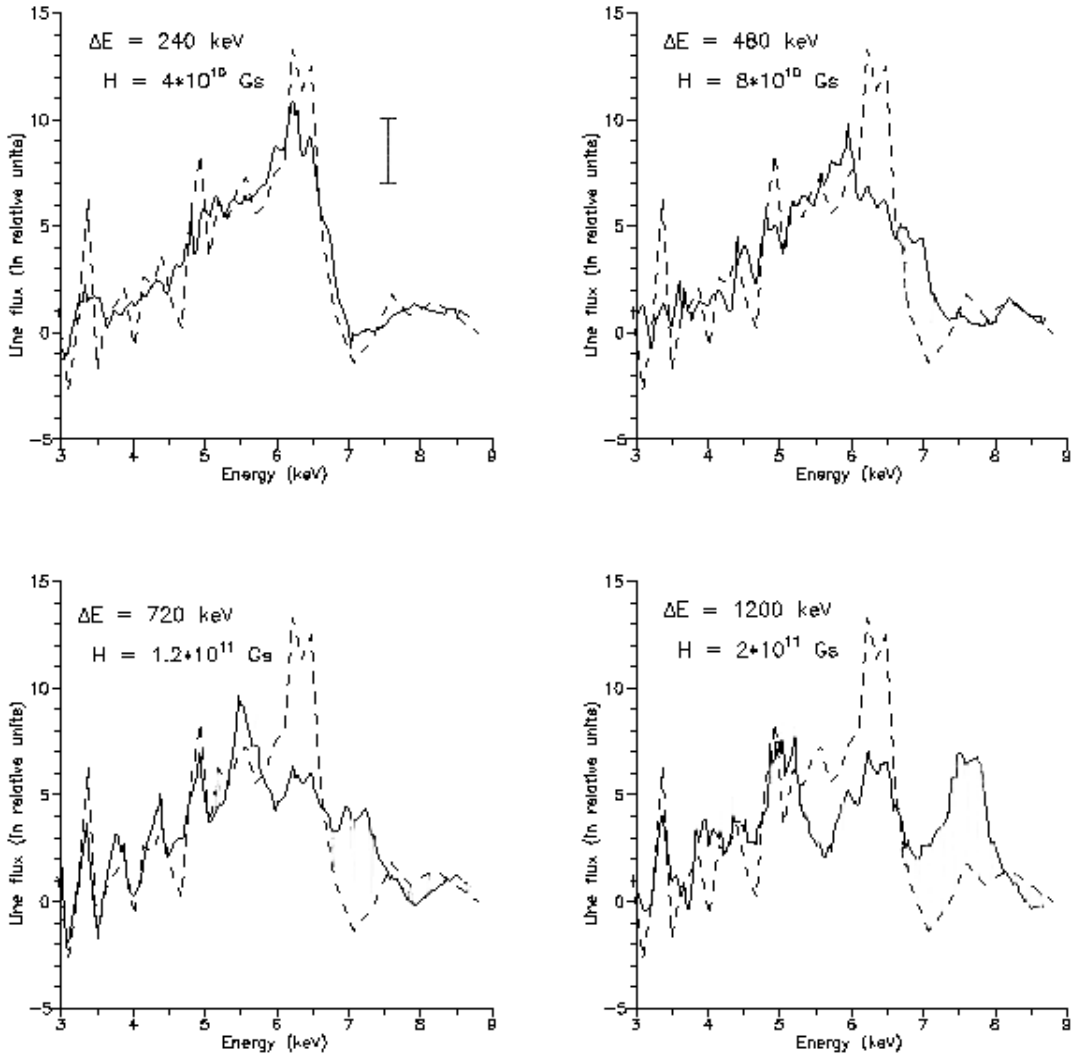,width=15cm}
  \caption{Influence of a magnetic field on the observational
           data. The dashed line represents the ASCA observations
           of MCG-6-30-15 \citep{tanaka1}. The vertical bar 
           in the top left panel corresponds to a typical error 
           in observation data. Solid lines show possible 
           profiles of $K_\alpha$ line in a presence of a magnetic 
           field. The field value and the appropriate Zeeman 
           splitting are indicated in each panel.}
  \label{zeeman05}
\end{figure*}

    The line profile without any magnetic field is presented
in~Fig.~\ref{zeeman01} for different values of disc
inclination angles: $\theta = 60^\circ$ and $\theta = 50^\circ$
respectively. Note, than at $\theta = 50^\circ$  the blue
peak appears more tall and narrow. Figs.~\ref{zeeman02},\ref{zeeman03} 
present the line profiles for the same inclination angles and 
different values of magnetic field: $H = 4\cdot 10^{10}$, 
$8\cdot 10^{10}$, $1.2\cdot 10^{11}$, $2\cdot 10^{11}$~Gs.
At $H = 4\cdot 10^{10}$~Gs the shape of spectral line
does not practically differ from the one with zero
magnetic field. Three components of the blue peak are so thin
and narrow that they could be scarcely distinguished 
experimentally today. For $H < 4\cdot 10^{10}$~Gs
and $\theta = 60\degr$ the splitting of the line
does not arise at all. At $\theta = 50\degr$ the splitting can 
still be revealed for $H = 3\cdot 10^{10}$~Gs 
(see Fig.~\ref{zeeman04}), but below this value 
($H < 3\cdot 10^{10}$~Gs) it also disappears.
With increasing the field the splitting becomes more explicit,
and at $H = 8\cdot 10^{10}$~Gs a faint hope appears to
register experimentally the complex internal structure of the
blue maximum. 
     
     While further increasing the magnetic field
the peak profile structure becomes apparent and can be distinctly
revealed, however, the field $H = 2\cdot 10^{11}$~Gs is rather
strong, so that the classical linear expression
for the Zeeman splitting
\begin{equation}
      \Delta E = \pm \mu_B H
      \label{eq15a}
\end{equation}
should be modified. Nevertheless, we use Eq.(\ref{eq15a}) for 
any value of the magnetic field, assuming that the qualitative 
picture of peak splitting remains correct, whereas for
$H = 2\cdot 10^{11}$~Gs the exact maximum positions may appear 
slightly different. If the Zeeman energy splitting
$\Delta E$ becomes of the order of $E$, the line splitting
due to magnetic fields is described in a more complicated
way. The discussion of this phenomenon is not a point of
this paper, our goal is to pay attention to the
qualitative features of this effect.

    Fig.~\ref{zeeman05} demonstrates a possible influence
of the Zeeman effect on observational data. As an illustration
we consider the observations
of iron $K_\alpha$ line which have been carried out by 
ASCA for the galaxy MCG-6-30-15. They are presented in
Fig.~\ref{zeeman05} in the dashed curve. Let us assume that
the actual magnetic field in these data is negligible. Then 
we can simulate the influence of the
Zeeman effect on the structure of observations and see if 
the simulated data (with a magnetic field) can be
distinguishable within the current accuracy of the observations.
The results of the simulated observation for the different 
values of magnetic field are shown in Fig.~\ref{zeeman05} 
in solid line. From these figures one can see that classical 
Zeeman splitting in three components, which can be revealed
experimentally today, changes qualitatively the line profiles
only for rather high magnetic field.
Something like this structure can be detected, e.g. for
$H = 1.2\cdot 10^{11}$~Gs, but the reliable recognition
of three peaks here is hardly possible.

    Apparently, it would be more correctly to solve the
inverse problem: to try to determine the magnetic field 
in the disc, assuming that the blue maximum is already split
due to the Zeeman effect. However, this problem includes too many 
additional factors, which can affect on the interpretation. 
Thus, beside of magnetic field the line width depends on the 
accretion disc model as well as on the structure of emitting 
regions. Such kind problems may become actual with a much better 
accuracy of observational data in comparison with their current 
state.

\section{Discussion}

    It is evident that the cause of duplication (triplication)
of a blue peak could be not only the influence of a magnetic 
field (the Zeeman effect), but a number of other factors. 
For example, the line profile can have two peaks when the emitting 
region represents two narrow rings with different values 
of radial coordinate (it is easy to conclude that two emitting 
rings with finite widths separated by a gap, would yield a similar 
effect). Despite of the fact that a multiple blue peak can originate 
from many causes (including the Zeeman effect as one 
of possible explanation), the absence of the multiple peak can 
result in the conclusion about the upper limit of the magnetic field.

    It is known that neutron stars (pulsars) could have
huge magnetic fields. So, it means that the discussed above
effect could appear in binary neutron star systems.
The quantitative description, however, needs more
detailed computations for such systems.

    With further increasing of observational facilities it
may become possible to improve the above estimation.
Thus, the Constellation-X launch suggested in the coming decade
seems to increase the precision of X-ray
spectroscopy as many as approximately 100 times with respect
to the present day measurements \citep{weaver1}.
Therefore, there is a possibility in principle
that the upper limit
of the magnetic field can also be 100 times improved in the case
when the emission of the X-ray line arise in a
sufficiently narrow region.

\section{Acknowledgements}

   One author (A.F.Z.) is grateful to E.F.Zakharova for
kindness and support, necessary to complete this work.

   S.V.R. is very grateful to Prof. E.V.Starostenko,
Dr. A.M.Salpagarov, Dr. O.N.Sumenkova and L.V.Bankalyuk
for the possibility of successful and intensive work over
the problem.

   The authors are grateful to E.A.Dorofeev, S.V.Molodtsov, L.Popovic and
A.I.Studenikin for useful discussion.

   This work has been partly supported by Russian
Foundation for Basic Research (grants 00--02--16108 (AFZ \& SVR),
01-02-16274 (VNL)) and the State Program in Astronomy (NSK).


\begin{thebibliography}{99}

\bibitem[\protect\citeauthoryear{Agol \& Krolik}{1999}]{Agol99}
   Agol E., Krolik J.H., 2000, ApJ, 528, 161 (astro-ph/9908049).

\bibitem[\protect\citeauthoryear{Asseo \& Sol}{1987}]{Ass87}
   Asseo E., Sol H.,
   1987, Physics Reports, 148, 307.

\bibitem[\protect\citeauthoryear{Ballantyne \& Fabian}{2001}]{Ball01}
   Ballantyne D.R., Fabian A.C.,
   2001, MNRAS,  328, L11 (astro-ph/0104342).

\bibitem[\protect\citeauthoryear{Ballantyne, Fabian \& Ross}{Ballantyne et al.}{2002}]{Ball02}
   Ballantyne D.R., Fabian A.C., Ross R.R.,
   2002, MNRAS, 329, L67 (astro-ph/0112179).

\bibitem[\protect\citeauthoryear{Balucinska-Church \& Church}{1999}]{Balu99}
   Balucinska-Church M., Church M.J.,
   2000, MNRAS, 312, L55 (astro-ph/9912389).

\bibitem[\protect\citeauthoryear{Berestetskii, Lifshits \& Pitaevskii}{Berestetskii et al.}{1982}]{BLP89}
   Berestetskii V.B., Lifshits E.M., Pitaevskii L.P., 1982,
   Quantum electrodynamics, Pergamon Press, Oxford.

\bibitem[\protect\citeauthoryear{Bianchi \& Matt}{2002}]{Bia02}
   Bianchi S., Matt G., 2002, A\&A, 387, 76.

\bibitem[\protect\citeauthoryear{Bisnovatyi-Kogan \& Ruzmaikin}{1974}]{Bis74}
   Bisnovatyi-Kogan G.S., Ruzmaikin A.A.,
   1974, A\&SS, 28, 45.

\bibitem[\protect\citeauthoryear{Bisnovatyi-Kogan \& Ruzmaikin}{1976}]{Bis76}
   Bisnovatyi-Kogan G.S., Ruzmaikin A.A.,
   1976, A\&SS, 42, 401.

\bibitem[\protect\citeauthoryear{Blokhintsev}{1963}]{Blokh63}
   Blokhintsev D.I., 1964, Quantum mechanics,
   D. Reidel Publ. Co. Dordrecht, Holland.

\bibitem[\protect\citeauthoryear{Boller et al.}{2001}]{boller01}
   Boller Th., Fabian A.C., Sunyaev R. et al.,
   2002, MNRAS, 329, L1 (astro-ph/0110367).

\bibitem[\protect\citeauthoryear{Bromley, Chen \& Miller}{Bromley et al.}{1997}]{bromley}
   Bromley B.C., Chen K., Miller W.A.,
   1997, ApJ, 475, 57.

\bibitem[\protect\citeauthoryear{Bromley, Miller \& Pariev}{Bromley et al.}{1998}]{bromley2}
   Bromley B.C., Miller W.A., Pariev V.I.,
   1998, Nature, 391, 54.

\bibitem[\protect\citeauthoryear{Carter}{1968}]{carter}
   Carter B., 1968,  Phys.Rev. D, 174, 1559.

\bibitem[\protect\citeauthoryear{Chandrasekhar}{1983}]{chandra}
   Chandrasekhar S., 1983, Mathematical Theory of Black Holes,
   Clarendon Press, Oxford.

\bibitem[\protect\citeauthoryear{Costa et al.}{2001}]{Cos01}
   Costa E., Soffitta P., Belazzini R. et al.,
   2001, Nature, 411, 662.

\bibitem[\protect\citeauthoryear{Cui, Zhang \& Chen}{Cui et al.}{1998}]{cui}
   Cui W., Zhang S.N., Chen W., 1998, ApJ, 257, 63.

\bibitem[\protect\citeauthoryear{Dirac}{1960}]{Dir60}
   Dirac P.A.M., 1958, The principles of quantum mechanics,
   4-th Edition, Oxford University Press.

\bibitem[\protect\citeauthoryear{Fabian et al.}{1995}]{fabian1}
   Fabian A.C., Nandra K., Reynolds C.S. et al.,
   1995, MNRAS, 277, L11.

\bibitem[\protect\citeauthoryear{Fabian}{1999}]{fabian99a}
   Fabian A.C., 1999, MNRAS, 308, L39 (astro-ph/9908064).

\bibitem[\protect\citeauthoryear{Fabian et al.}{2001}]{fabian2}
   Fabian A.C., 2001, in  Relativistic Astropysics,
   Texas Symposium, American Institute of Physics,
   AIP Conference Proceedings, 586, 643.

\bibitem[\protect\citeauthoryear{Fabian, Vaughan \& Nandra}{Fabian et al.}{2002}]{Fabian02}
   Fabian A.C., Vaughan S., Nandra K., accepted to MNRAS
   (astro-ph/0206095).

\bibitem[\protect\citeauthoryear{Fock }{1976}]{Fock76}
   Fock V.A., 1978, Fundamentals of quantum mechanics,
   Mir, Moscow.

\bibitem[\protect\citeauthoryear{Gear}{1971}]{gear}
   Gear C.W., 1971, Numerical Initial Value Problems in Ordinary
   Differential Equations.  Prentice Hall, Englewood Cliffs, NY.

\bibitem[\protect\citeauthoryear{Giovannini}{2001}]{Giov01}
   Giovannini M., hep-ph/0111220.

\bibitem[\protect\citeauthoryear{Greiner}{1999}]{grein99}
   Greiner J., 2000, in "Cosmic Explosions", Proceedings of 10th
   Annual Astrophysical Conference in Maryland, eds. S.Holt \&
   W.W.Zang, AIP Conference proceedings, 522, 307 (astro-ph/9912326).

\bibitem[\protect\citeauthoryear{Hiebert \& Shampine}{1983}]{hiebert}
   Hiebert K.L., Shampine L.F., 1980, Implicitly Defined Output
   Points for Solutions of ODE-s. Sandia report sand80--0180,
   February.

\bibitem[\protect\citeauthoryear{Hindmarsh}{1983}]{hindmarsh}
   Hindmarsh A.C., in Stepleman, R.S. et al. eds,
   ODEpack, a systematized collection of ODE solvers.
   In Scientific Computing, North--Holland, Amsterdam, p. 55.

\bibitem[\protect\citeauthoryear{Iwasawa et al.}{1999}]{Iwa99}
   Iwasawa K., Fabian A.C., Young A.J. et al.,
   1999, MNRAS, 306, L19 (astro-ph/9904078).

\bibitem[\protect\citeauthoryear{Karas, Martocchia \& Subr}{Karas et al.}{2001}]{karas01}
   Karas V., Martocchia A., Subr L.,
   2001, PASJ, 53, 189 (astro-ph/0102460).

\bibitem[\protect\citeauthoryear{Kardashev}{1995}]{Kard95}
   Kardashev N.S., 1995,  MNRAS, 276, 515.

\bibitem[\protect\citeauthoryear{Kardashev}{2001a}]{Kard00}
   Kardashev N.S., 2001a, in Giancarlo Setti and Jean-Pierre Swings eds.,
   Quasars, AGNs and Related Research Across 2000.
   Conference on the occasion of L.Woltjer's 70th birthday, Rome,
   Italy, May 2000. Springer, p.66.

\bibitem[\protect\citeauthoryear{Kardashev}{2001b}]{Kard01a}
   Kardashev N.S., 2001b, in Kardashev N.S., Dagkesamanskij R.D.,
   Kovalev Yu.A. eds, Astrophysics on the edge of centuries.
   Proceedings of Russian astronomical conference, Pushchino.
   Yanus-K, Moscow, p. 383.

\bibitem[\protect\citeauthoryear{Kardashev}{2001c}]{Kard01}
   Kardashev N.S., 2001c,  MNRAS, 326, 1122.

\bibitem[\protect\citeauthoryear{Koide et al.}{2002}]{Koide02a}
   Koide S., Shibata K., Kudoh T. et al., 2002, Science, 295, 1688.

\bibitem[\protect\citeauthoryear{Krolik}{1999}]{Krolik99}
   Krolik J.H., 1999, ApJ, 515, L73 (astro-ph/9902267).

\bibitem[\protect\citeauthoryear{Krolik}{2001}]{krolik01}
   Krolik J.H., 2001, in  Relativistic Astropysics,
   Texas Symposium, American Institute of Physics,
   AIP Conference Proceedings, 586, p. 674.

\bibitem[\protect\citeauthoryear{Landau \& Lifshits}{1975}]{fieldtheory}
   Landau L.D., Lifshits E.M.,
   The classical theory of fields,
   1975, Pergamon Press, Oxford.

\bibitem[\protect\citeauthoryear{Lazzati et al.}{2001}]{Lazz01}
   Lazzati D., Ghisellini G., Vietri M. et al., 2001,
   in Enrico Costa, Filippo Frontera, Jens Hjorth eds,
   Proceedings of the International workshop held in Rome,
   CNR headquaters. Springer, Berlin Heidelberg, p. 236
   (astro-ph/0104086).

\bibitem[\protect\citeauthoryear{Lee et al.}{1999}]{Lee99}
   Lee J.C., Fabian A.C., Reynolds C.S. et al.,
   2000, MNRAS, 318, 857 (astro-ph/9909239).

\bibitem[\protect\citeauthoryear{Lee et al.}{2002}]{Lee02}
   Lee J.C., Iwasawa K., Houck J.C. et al.,
   2002, ApJ, 570, L47 (astro-ph/0203523).

\bibitem[\protect\citeauthoryear{Levenson et al.}{2002a}]{Lev02}
   Levenson N.A. et al.,
   2002, ApJ, 573, L84.

\bibitem[\protect\citeauthoryear{Levenson et al.}{2002b}]{Lev02a}
   Levenson N.A., Krolik J.H., Zycki P.T. et al.,
   2002, ApJ, 573, L81 (astro-ph/0206071).

\bibitem[\protect\citeauthoryear{Liang}{1998}]{Liang98}
   Liang E.P.,    1998, Physics Reports, 302, 69.

\bibitem[\protect\citeauthoryear{Lipunova \& Shakura}{2002}]{LipShak}
   Lipunova G.V., Shakura N.I.,
   2002, Astronomy Reports, 46, 366.

\bibitem[\protect\citeauthoryear{Lovelace et al.}{1998}]{romanova2}
   Lovelace R.V.E., Newman W.I., Romanova M.M.,
   1997, ApJ, 484, 628.

\bibitem[\protect\citeauthoryear{Lovelace et al.}{1999}]{Love99}
   Lovelace R.V.E., Li H., Colgate S.A. et al.,
   1999, ApJ, 513, 805.

\bibitem[\protect\citeauthoryear{Malizia et al.}{1997}]{malizia}
   Malizia A., Bassani L., Stephen J.B. et al.,
   1997, ApJSS, 113, 311.

\bibitem[\protect\citeauthoryear{Martocchia et al.}{2002a}]{Mart02a}
   Martocchia A., Matt G., Karas V. et al.,
   2002, A\&A, 387, 215 (astro-ph/0203185).

\bibitem[\protect\citeauthoryear{Ma}{2000}]{Ma00}
   Ma Zhen-guo,  2000, Chinese A\&A, 24, 135.

\bibitem[\protect\citeauthoryear{Ma}{2002}]{Ma02}
   Ma Zhen-guo,  2002,
   in Proc. IAU 214 Symposium (in press).

\bibitem[\protect\citeauthoryear{Martocchia, Matt \& Karas}{Martocchia et al.}{2002b}]{Mart02}
   Martocchia A., Matt G., Karas V.,
   2002, A\&A, 383, L23 (astro-ph/0201192).

\bibitem[\protect\citeauthoryear{Paul et al.}{1992}]{matt1}
   Matt G., Perola G.C., Piro L., Stella L.,
   1992, A\&A, 475, 57.

\bibitem[\protect\citeauthoryear{Matt}{2002}]{matt02}
   Matt G.,
   accepted in MNRAS (astro-ph/0207615).

\bibitem[\protect\citeauthoryear{Meier, Koide \& Uchida}{2001}]{Koide02}
   Meier D.L., Koide S., Uchida Y., 2001, Science, 291, 84.

\bibitem[\protect\citeauthoryear{Messia}{1976}]{Mess79}
   Messia A., 1976, Quantenmechanik, Walter de Geuter,
   Berlin.

\bibitem[\protect\citeauthoryear{Miller et al.}{2002}]{Miller02}
   Miller J.M., Fabian A.C., Wijnands R. et al.,
   2002, ApJ, 570, L73.

\bibitem[\protect\citeauthoryear{Mirabel}{2000}]{mira00}
   Mirabel I.F, 2001, A\&SSS, 276, 319 (astro-ph/0011315).

\bibitem[\protect\citeauthoryear{Mirabel}{2002}]{Mira02}
   Mirabel I.F, 2002, private communication.

\bibitem[\protect\citeauthoryear{Mirabel \& Rodriguez}{2002}]{Mira02a}
   Mirabel I.F, Rodriguez L.F., 2002, Sky \& Telescope, May, 33.

\bibitem[\protect\citeauthoryear{Misner et al.}{1973}]{wheeler}
   Misner C.W., Thorne K.S., Wheeler J.A., 1973, Gravitation.
   W.H.Freeman, San Francisco.

\bibitem[\protect\citeauthoryear{Morales \& Fabian}{2001}]{moral01}
   Morales R., Fabian A.C.,
   2002, MNRAS, 329, 209 (astro-ph/0109050).

\bibitem[\protect\citeauthoryear{Nandra et al.}{1997a}]{nandra1}
   Nandra K., George I.M., Mushotzky R.F. et al.,
   1997, ApJ, 476, 70.

\bibitem[\protect\citeauthoryear{Nandra et al.}{1997b}]{nandra2}
   Nandra K., George I.M., Mushotzky R.F. et al.,
   1997, ApJ, 477, 602.

\bibitem[\protect\citeauthoryear{Novikov \& Frolov}{2001}]{FN01}
   Novikov I.D., Frolov V.P., 2001, Physics -- Uspekhi, 44,
   291.

\bibitem[\protect\citeauthoryear{Novikov \& Thorne}{1973}]{novikov}
   Novikov I.D., Thorne K.S., 1973,
   in De Witt C., De Witt B.S. eds, Black Holes.
   New York: Gordon \& Breach, p. 334.

\bibitem[\protect\citeauthoryear{Ogle et al.}{2000}]{ogle1}
   Ogle, P.M., Marshall, H.L., Lee J.C. et al.,
   2000, ApJ, 545, L81.

\bibitem[\protect\citeauthoryear{Pariev \& Bromley}{1997}]{pariev2}
   Pariev V.I, Bromley B.C., 1997, Proceedings of the 8-th Annual October
   Astrophysics Conference in Maryland (astro-ph/9711214).

\bibitem[\protect\citeauthoryear{Pariev \& Bromley}{1998}]{pariev1}
   Pariev V.I., Bromley B.C., 1998, ApJ, 508, 590.

\bibitem[\protect\citeauthoryear{Pariev, Bromley \& Miller}{Pariev et al.}{2000}]{pariev3}
   Pariev V.I., Bromley B.C., Miller W.A.,
   2001, ApJ, 547, 649 (astro-ph/0010318).

\bibitem[\protect\citeauthoryear{Paul et al.}{1998}]{paul}
   Paul, B., Agrawal, P.C., Rao, A.R. et al.,
   1998, ApJ, 492, 15.

\bibitem[\protect\citeauthoryear{Petzold}{1983}]{petzold}
   Petzold L.R., 1983, SIAM J. Sci. Stat. Comput, 4, 136.

\bibitem[\protect\citeauthoryear{Popovic, Mediavilla \& Munoz}{Popovic et al.}{2001}]{Popov01}
   Popovi\'c L.C., Mediavilla E.G., Munoz J.A.,
   2001, A\&A, 378, 295.

\bibitem[\protect\citeauthoryear{Popovic et al.}{2002}]{Popov02}
   Popovi\'c L.C., Mediavilla E.G., Jovanovi\'c P. et al.
   2002, accepted


\bibitem[\protect\citeauthoryear{Qingjuan \& Youjun}{2001}]{Qing01}
   Qingjuan Yu, Youjun Lu,
   2001, A\&A, 377, 17.

\bibitem[\protect\citeauthoryear{Romanova et al.}{1998}]{romanova1}
   Romanova M.M., Ustyugova G.V., Koldoba A.V. et al.,
   1998, ApJ, 500, 703.

\bibitem[\protect\citeauthoryear{Sambruna et al.}{1998}]{sambruna}
   Sambruna R.M., George I.M., Mushotsky R.F. et al.,
   1998, ApJ, 495, 749.

\bibitem[\protect\citeauthoryear{Sazonov, Churazov \& Sunyaev}{Sazonov et al.}{2002}]{Saz02}
   Sazonov S.Yu., Churazov E.M., Sunyaev R.A.,
   2002, MNRAS, 330, 817.

\bibitem[\protect\citeauthoryear{Shakura}{1973}]{shakura}
   Shakura N.I., 1972, A. Zh., 49, 921.

\bibitem[\protect\citeauthoryear{Shakura \& Sunyaev}{1973}]{shasun}
   Shakura N.I., Sunyaev R.A., 1973, A\&A, 24, 337.

\bibitem[\protect\citeauthoryear{Shih, Iwasawa \& Fabian}{Shih et al.}{2002}]{Shih02}
   Shih D.C., Iwasawa K., Fabian A.C.,
   2002, MNRAS, 2002, 333, 687 (astro-ph/9904078).

\bibitem[\protect\citeauthoryear{Sulentic et al.}{1998b}]{sulentic2}
   Sulentic J.W., Marziani P., Zwitter T. et al.,
   1998, ApJ, 501, 54.

\bibitem[\protect\citeauthoryear{Sulentic et al.}{1998a}]{sulentic1}
   Sulentic J.W., Marziani P., Calvani M.,
   1998, ApJ, 497, L65.

\bibitem[\protect\citeauthoryear{Tanaka et al.}{1995}]{tanaka1}
   Tanaka Y., Nandra K., Fabian A.C. et al.,
   1995, Nature, 375, 659.

\bibitem[\protect\citeauthoryear{Turner et al.}{2002}]{Tur02}
   Turner T. J., Mushotzky R.F., Yagoob T. et al.,
   2002, ApJ, 574, L127.

\bibitem[\protect\citeauthoryear{Vietri \& Stella}{1998}]{vietri98}
   Vietri M.,  Stella L.,
   1998, ApJ, 503,350 (astro-ph/9803089).

\bibitem[\protect\citeauthoryear{Wanders et al.}{1997}]{wanders}
   Wanders I. et al.,
   1997, ApJSS, 113, 69.

\bibitem[\protect\citeauthoryear{Weawer, Krolik \& Pier}{Weawer et al.}{1998}]{krolik1}
   Weawer K.A., Krolik J.H., Pier E.A.,
   1998, ApJ, 498, 213, (astro-ph/9712035).

\bibitem[\protect\citeauthoryear{Weawer}{2001}]{weaver1}
   Weaver K.A., 2001, in  Relativistic Astropysics,
   Texas Symposium, American Institute of Physics,
   AIP Conference Proceedings, 586, p. 702.

\bibitem[\protect\citeauthoryear{Wilms, Reynolds \& Begelman}{Wilms et al.}{2001}]{Wilms01}
   Wilms J., Reynolds C.S., Begelman M.C. et al.
   2001, MNRAS, 328, L27 (astro-ph/0110520).

\bibitem[\protect\citeauthoryear{Yaqoob et al.}{1996}]{yaqoob1}
   Yaqoob T., Serlemitsos P.J., Turner T.J. et al.,
   1996, ApJ, 470, L27.

\bibitem[\protect\citeauthoryear{Yaqoob et al.}{1997}]{yaqoob2}
   Yaqoob T., McKernan B., Ptak A. et al.,
   1997, ApJ, 490, L25.

\bibitem[\protect\citeauthoryear{Yaqoob, George \& Turner}{Yaqoob et al.}{2001a}]{yaqoob01a}
   Yaqoob T., George I.M., Turner T.J., astro-ph/0111428.

\bibitem[\protect\citeauthoryear{Yaqoob et al.}{2001b}]{yacoob4}
   Yaqoob T., George I.M., Nandra, K. et al.,
   2001, ApJ, 546, 759.

\bibitem[\protect\citeauthoryear{Yaqoob et al.}{2002}]{yaqoob02}
   Yaqoob T., Padmanabhan U., Dotani T. et al.,
   2002, ApJ, 569, 487.

\bibitem[\protect\citeauthoryear{Zakharov}{1986}]{zakh86}
   Zakharov A.F., 1986, JEThP, 91, 3.

\bibitem[\protect\citeauthoryear{Zakharov}{1989}]{zakh89}
   Zakharov A.F., 1989, JEThP, 95, 385.

\bibitem[\protect\citeauthoryear{Zakharov}{1991}]{zakh91}
   Zakharov A.F., 1991, Soviet Astronomy, 35, 147.

\bibitem[\protect\citeauthoryear{Zakharov}{1993}]{zakharov6}
   Zakharov A.F., 1993, Preprint MPA 755.

\bibitem[\protect\citeauthoryear{Zakharov}{1994}]{zakharov1}
   Zakharov A.F., 1994, MNRAS, 269, 283.

\bibitem[\protect\citeauthoryear{Zakharov}{1995}]{zakharov5}
   Zakharov A.F., 1995, in Annals for the 17th Texas Symposium
   on Relativistic Astropysics, The New York Academy of Sciences,
   759, 550.

\bibitem[\protect\citeauthoryear{Zakharov}{2000}]{Zak00}
   Zakharov A.F., 2000, in Proc. of the XXIII Workshop
   on High Energy Physics and Field Theory, IHEP, Protvino,
   169.

\bibitem[\protect\citeauthoryear{Zakharov \& Repin }{1999}]{zak_rep1}
   Zakharov A.F., Repin S.V., 1999, Astronomy Reports, 43, 705.

\bibitem[\protect\citeauthoryear{Zakharov \& Repin }{2002a}]{zak_rep2}
   Zakharov A.F., Repin S.V., 2002, Astronomy Reports, 46, 360.

\bibitem[\protect\citeauthoryear{Zakharov \& Repin }{2002b}]{zak_rept}
   Zakharov A.F., Repin S.V., 2002,
in J. Koga, T. Nakamura, K. Maeda, K. Tomita (eds), Proc. of the Eleven  Workshop
   on General Relativity  and Gravitation in Japan, Waseda University, Tokyo,
p.~ 68.


\bibitem[\protect\citeauthoryear{Zamanov \& Marziani}{2002}]{zaman02}
   Zamanov R., Marziani P., astro-ph/0204423 (accepted in ApJ).








%
%
%
%
%
%
%
%

\end{thebibliography}
\end{document}